\documentclass[%
 reprint,
superscriptaddress,
showkeys,
 amsmath,amssymb,
 aps,
pre, 
]{revtex4-2}

\usepackage{graphicx}% Include figure files
\usepackage[caption=false]{subfig} %?

\usepackage{dcolumn}% Align table columns on decimal point
\usepackage{bm}% bold math
\newcommand{\ee}{\end{equation}} 
\newcommand{\be}{\begin{equation}}

\usepackage[mathscr]{euscript}  
\usepackage{xcolor}  

\usepackage{tcolorbox}
\usepackage{comment}

\makeatletter
\newsavebox{\@brx}
\newcommand{\llangle}[1][]{\savebox{\@brx}{\(\m@th{#1\langle}\)}%
  \mathopen{\copy\@brx\kern-0.5\wd\@brx\usebox{\@brx}}}
\newcommand{\rrangle}[1][]{\savebox{\@brx}{\(\m@th{#1\rangle}\)}%
  \mathclose{\copy\@brx\kern-0.5\wd\@brx\usebox{\@brx}}}
\makeatother

\begin{document}

\preprint{APS/123-QED}
%\title{ Perturbative effective diffusivity of microswimmers\\ in the presence of oscillating torques }

\title{ Diffusion of active particles with angular velocity reversal}

\author{Kristian St\o{}levik Olsen}
\affiliation{PoreLab, Department of Physics, University of Oslo, Blindern, 0316 Oslo, Norway\\}

\date{\today}

\begin{abstract}
Biological and synthetic microswimmers display a wide range of swimming trajectories depending on driving forces and torques. In this paper we consider a simple overdamped model of self-propelled particles with a constant self-propulsion speed, but an angular velocity that varies in time. Specifically, we consider the case of both deterministic and stochastic angular velocity reversal, mimicking several synthetic active matter systems like propelled droplets. The orientational correlation function and effective diffusivity is studied using Langevin dynamics simulations and perturbative methods. 
\end{abstract}

\pacs{Valid PACS appear here} % PACS, the Physics and Astronomy  Classification Scheme.

\maketitle

%%%%%%%%%%%%%%%%%%%%%%%%%%
\section{Introduction}\label{sec:int}
In the past decades the field of active matter has grown substantially both in interest and application \cite{marchetti2013hydrodynamics,needleman2017active, gompper20202020}. Ranging from bio-inspired micro- and nano-robotics and engines to crowd behavior, the applications of the ideas in active matter research spans a multitude of length scales \cite{pincce2016disorder, pietzonka2019autonomous, palagi2018bioinspired,  di2010bacterial, moreno2020collective, kulkarni2019sparse}. Particular focus perhaps has been devoted to relatively simple theoretical models mimicking the behavior of living systems, where aspects such as self-propulsion are central \cite{ebeling1999active, bernheim2018living, fodor2018statistical, zhou2014living}. Such systems generally break time-reversal symmetry, by energy being injected locally and then dissipated, and are driven out of equilibrium. Chiral active matter presents a relatively new class of non-equilibrium systems, where not only energy is injected on the particle scale but also angular momentum \cite{lowen2016chirality}. The prototypical example of such behavior is that of bacterial motion in the presence of walls or boundaries, where chiral trajectories with a given handedness is observed \cite{lauga2006swimming,park2019flagellated}.

Synthetic active matter systems have also gained a lot of interest is recent times, both due to the relatively simple experimental setups that reveals fascinating non-equilibrium phenomena, and because of the possible applications in medicine and drug delivery. Several investigations have for example looked into the possibility of sorting and separating active particles of differing chiralities, with important applications in the pharmaceutical industry \cite{lin2020separation,chen2015sorting,levis2019activity, mijalkov2013sorting}. Recent studies have revealed a myriad of swimming path shapes for synthetic microswimmers, with examples including zig-zag motion of swimming droplets driven by mechanical agitation \cite{ebata2015swimming}, and the motion of deformable self-propelled particles under forcing \cite{tarama2011dynamics}. Likewise, nematic liquid droplets have also shown swimming paths where the handedness of the droplets alternate in time, here due to autochemotactic reorientations \cite{hokmabad2019topological, kruger2016curling}. 

It is our intention in this paper to theoretically investigate the transport properties of active particles that exhibit angular velocity reversal, which we model by including either deterministic or stochastic terms in the angular equations of motion. We use a simple overdamped model of chiral active Brownian point-particles, as described below. While the transport properties of chiral active matter has been studied analytically in the past \cite{PhysRevE.94.062120,caprini2019active}, the case of time-dependent angular velocities has received less attention. Some studies consider the effect of continuous fluctuations around a mean chirality \cite{weber2011active,notel2016diffusion}, originating for example from an imperfect external driving or from internal processes in biological applications, while we here focus on angular dynamics mimicking the zig-zag motion of particles that reverse their angular velocity either in a predictable manner or through sudden stochastic reversal events.  We report on analytical formulas for the effective late-time diffusivity which we calculate perturbatively in the regime where switching frequency is large compared to the maximal value of the angular velocity. Results are verified using Langevin dynamics simulations, and the regime outside of the perturbative analysis is also investigated numerically.  While time-dependent angular dynamics has been observed also in three dimensions \cite{tjhung2017contractile}, we here restrict our attention to a minimal two-dimensional model. Theoretical methods proposed in the past for dealing with three-dimensional chiral motion, eg. that of Sevilla \cite{PhysRevE.94.062120}, could potentially be generalized to also non-trivial angular dynamics, which presents an intriguing extension of this paper. Such methods could also give further insights into spatial distributions of active particles, while the Green-Kubo style framework utilized here allow us to easily estimate moments and correlation functions.

\begin{figure}[t]
    \centering
    \includegraphics[width = 8.0cm]{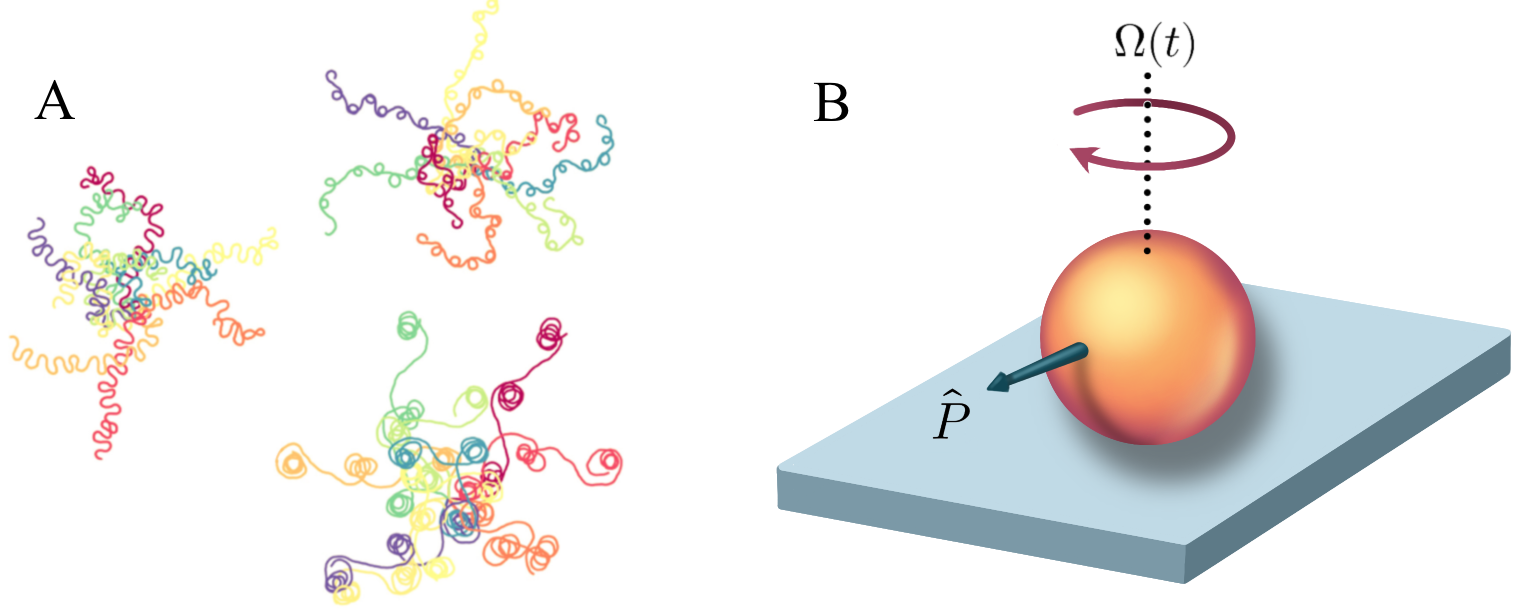}
    \caption{A) Typical trajectories for chiral particles with oscillating torque displaying both zig-zag/meandering and curling behavior, corresponding to the model in section III. B) Chiral particles with a direction of motion given by the vector $\hat P(\phi)$ moving in the plane with an angular velocity $\Omega(t)$.}
    \label{fig:traj}
\end{figure}

The paper is structured as follows. Section II discusses the general theoretical setup of the model considered in this paper, where particles in addition to experiencing angular noise has a non-trivial deterministic angular dynamics. Section III considers the case of a deterministic angular velocity reversal, which we model using a sinusoidal time dependence. The correlation function of the direction of motion is calculated exactly, and the effective diffusivity is studied perturbatively. Section IV considers a related model where the reversal is stochastic. Predictions of an effective persistence time and diffusivity in a perturbative regime is given and numerically verified. A concluding discussion is offered in section V.

\begin{figure}[t]
    \centering
    \includegraphics[width = 8.6cm]{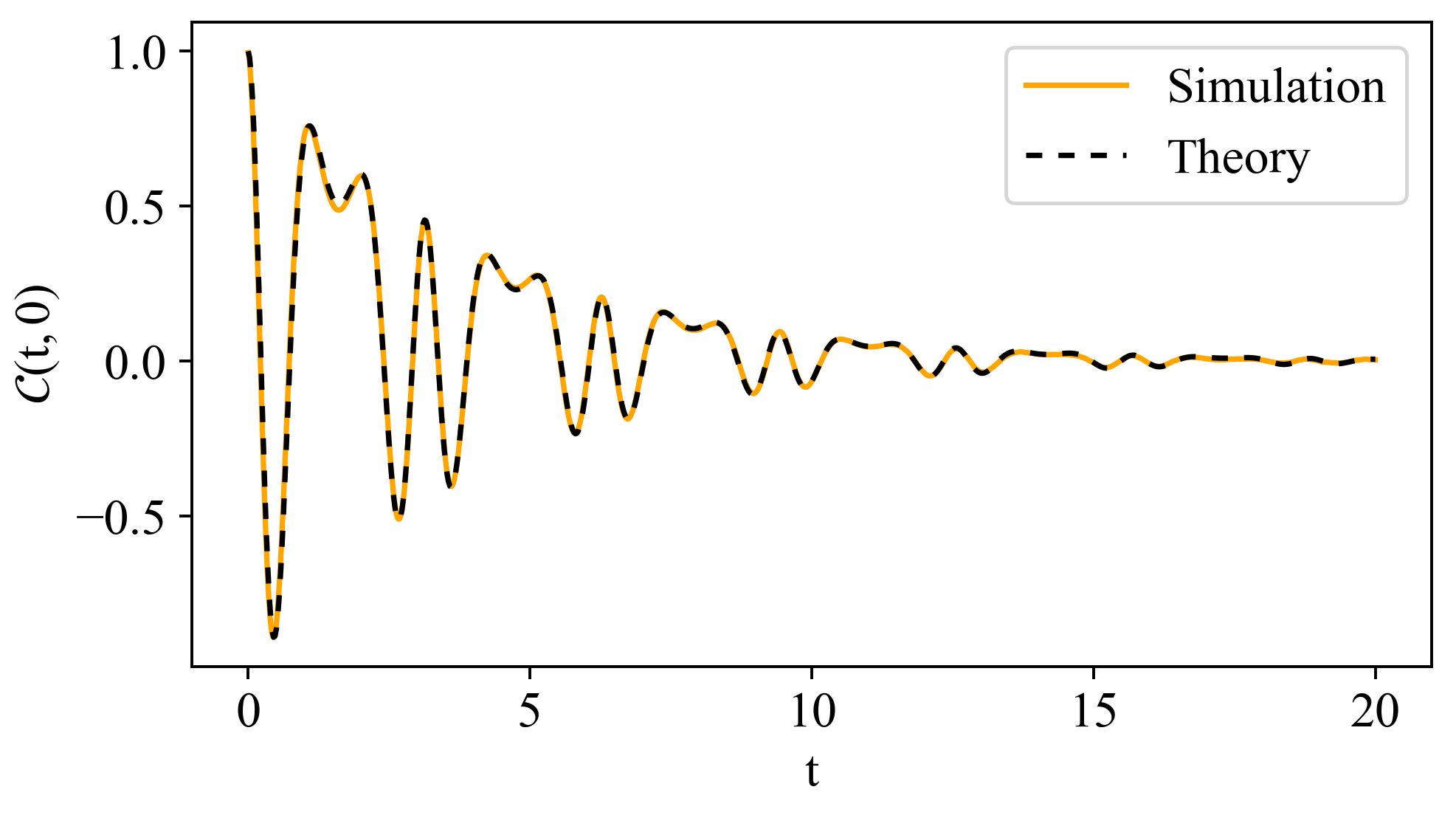}
    \caption{Correlation function for direction of motion for active particle with deterministic angular velocity reversal. Solid line resulting from ensemble average over $N = 5\cdot 10^4$ particle trajectories, and dashed line showing Eq. (\ref{eq:corr_deterministic}). The enveloping exponential decays with the standard persistence time scale $1/D_\phi$. Parameters used: $D_\phi = 0.25, \Omega_0 = 7,\gamma = 1$. }
    \label{fig:corr_det}
\end{figure}

%%%%%%%%%%%%%%%%%%%%%%%%%%
\section{Active Brownian particles with time-dependent torques}\label{sec:theory}

We consider self-propelled particles constrained to two dimensions described by overdamped Langevin equations. The dynamics follows 

\begin{align}
    & \dot x_\alpha(t) = u_0 \hat P_\alpha(\phi),  \label{eq:L1} \\
    &\dot \phi(t) = \sqrt{2 D_\phi}\zeta(t) + \Omega(t), \label{eq:L2}
\end{align}
where $\hat P = (\cos\phi,\sin\phi)$ is the unit vector pointing in the direction of motion and $u_0$ is the constant self-propulsion speed of the particles. Here we used index notation with $\alpha$ a spatial index. The diffusion coefficient $D_\phi$ governs the angular noise, which sets the persistence time-scale $\tau_0 = 1/D_\phi$ for the particle changing its direction of motion due to noise alone. The noise is Gaussian and white with $\langle \zeta(t)\rangle= 0$ and $\langle \zeta(t_1)\zeta(t_2)\rangle= \delta(t_1-t_2)$.

The term $\Omega(t)$ is the time-dependent angular velocity of the particles, originating from a time-dependent torque (see fig. \ref{fig:traj}). In the case of a constant $\Omega$ this model is the simplest model of chiral active Brownian particles that perform circular trajectories, while the case $\Omega = 0$ corresponds to linear swimmers.

The class of angular velocities considered here give rise to behaviors that are neither chiral nor linear. Much information regarding the late-time dynamics can be extracted from the correlation function for the direction of motion, defined as 
\begin{equation}
    \mathcal{C}(t_2,t_1) = \langle \hat P(t_1) \cdot \hat P(t_2)\rangle =  \langle \cos[\phi(t_2)-\phi(t_1)]\rangle.
\end{equation}
This correlation function can readily be calculated from the Langevin equations by using the fact that the angular dynamics in integral form can be writen 
\begin{equation}
    \phi(t) = \sqrt{2 D_\phi} W(t) + \int_0^t ds \Omega(s)
\end{equation}
where $W$ is the Wiener process. Using standard properties of the Wiener process, we may write the difference in angles as
\begin{equation}
    \phi(t_2)- \phi(t_1) = \sqrt{2 D_\phi (t_2-t_1)} Z +\int_{t_1}^{t_2} ds \Omega(s)
\end{equation}
for $t_1<t_2$, where $Z$ is an independent normal random variable with unit variance. Using standard trigonometric identities we may then write 
\begin{align}
     \mathcal{C}(t_2,t_1)  =   &\left\langle\cos \left[ \sqrt{2 D_\phi (t_2-t_1)} Z \right]\right\rangle  \cos \int_{t_1}^{t_2} ds \Omega(s)
\end{align}
where the sinusoidal contribution vanishes from symmetry arguments. The expectation value may be calculated independently of the deterministic contribution from the angular velocity. We have 
\begin{equation}
    \left\langle\cos \left[ \sqrt{2 D_\phi (t_2-t_1)} Z \right]\right\rangle = e^{-D_\phi(t_2-t_1)}
\end{equation}
  which may for example be seen by expressing the cosine as an infinite series, in which case the moments of $Z$ can be estimated and the series re-summed. Hence the correlation function takes the form 
\begin{equation}\label{eq:corr}
    \mathcal{C}(t_2,t_1) = e^{-D_\phi (t_2-t_1)}\cos \int_{t_1}^{t_2} ds \Omega(s), t_2>t_1.
\end{equation}
This reduces to the well known correlation function for the direction of motion of active Brownian particles in the case of constant $\Omega$. Note the integration inside the cosine which breaks time-translational symmetry.

\begin{figure*}[t]
    \centering
    \includegraphics[width = 17.7cm]{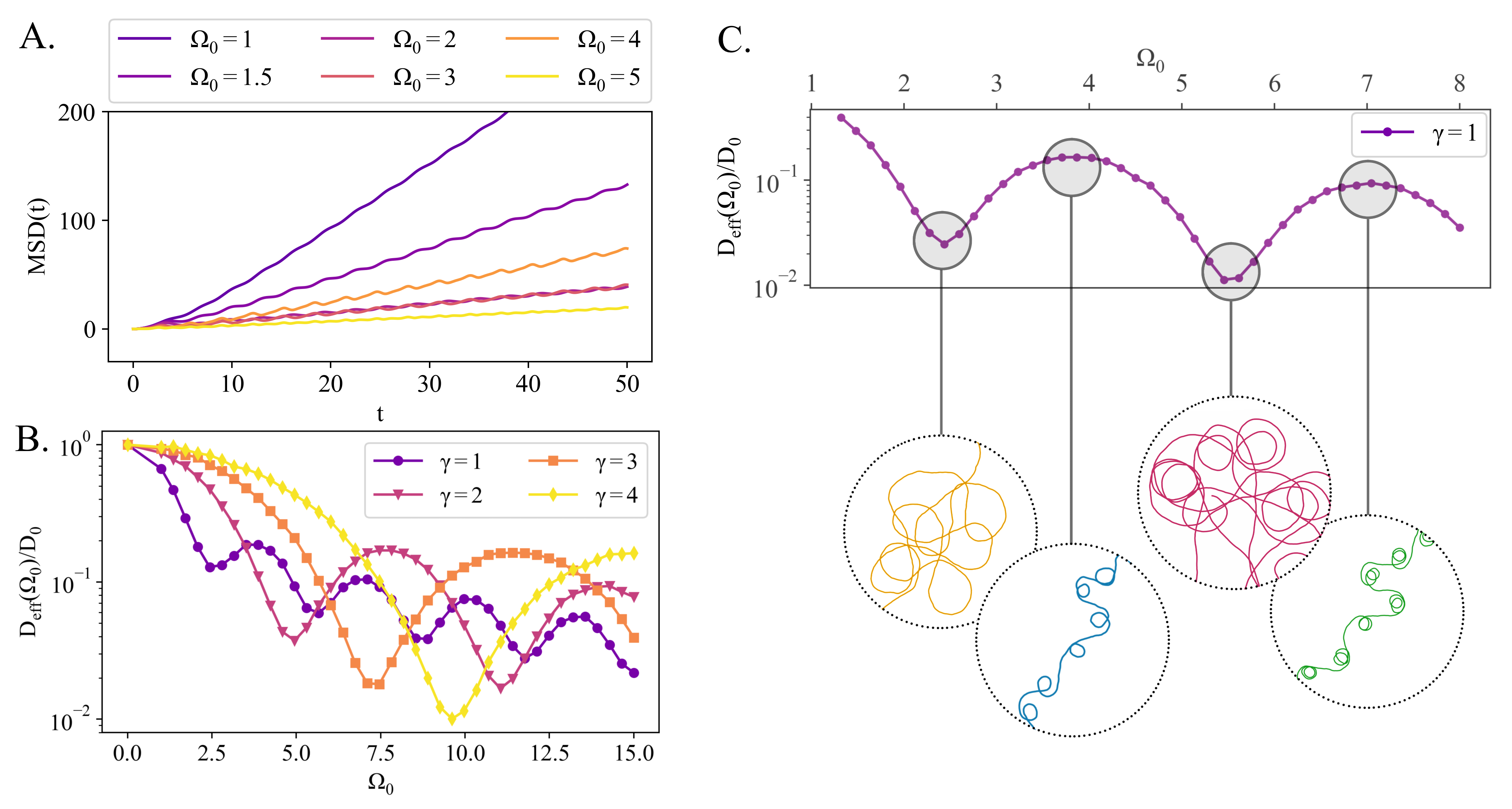}
    \caption{Figure showing mean-squared displacements (A) and the associated effective diffusivity (B), both obtained numerically by simulating a large number of independent active particles using the equations of motion. A non-monotonic behavior in the effective diffusivity is observed, stemming from a competition between the two time scales set by the maximal angular speed $\Omega_0$ and the reversal rate $\gamma$. Part C of the figure shows example trajectories corresponding to peaks and valleys in the plot of the effective diffusivity.   Parameters used are $D_\phi = 0.2, \Omega_0 = 7, u_0 = 1, \gamma = 1$  unless otherwise stated in the figures.}
    \label{fig:collected}
\end{figure*}

At late times we expect the mean-square displacement to scale as $\langle\Delta \vec{x}^2(t)\rangle = 2 D_\text{eff} t$, with the effective diffusivity taking the form 
\begin{equation}\label{eq:first}
    D_\text{eff} = u_0^2 \lim_{t\to \infty} \partial_t \int_0^{t} dt_2 \int_0^{t_2} dt_1 \mathcal{C}(t_1,t_2),
\end{equation}
similar to expressions derived in the past when combined with Eq. (\ref{eq:corr}) \cite{hagen2009non}. In the case of a constant angular velocity, $\Omega_0$, this equation results in the known result for chiral active Brownian particles \cite{lowen2020inertial}
\begin{equation}
D_\text{eff} = \frac{u_0^2}{D_\phi} \frac{1}{1 + \left(\frac{\Omega_0}{D_\phi}\right)^2}.
\end{equation}
As can be seen, chirality suppresses transport. For the case of a linear swimmer ($\Omega_0 =0$) we denote the diffusivity as $D_0 = u_0^2/D_\phi$. This will act as a reference diffusivity to which other effective diffusivities are compared.

%Fig. (\ref{fig:traj}A) shows typical trajectories in the low noise regime, obtained by numerically integrating the above stochastic dynamics. 

\section{Deterministic angular velocity reversal}

For the case of deterministic angular motion reversal, we consider the angular velocity $\Omega(t) = \Omega_0 \cos(\gamma t)$ where $\Omega_0$ is the magnitude of angular velocity and $\gamma$ the reversal rate. The correlation function can in this case be written 
\begin{equation}\label{eq:corr_deterministic}
    \mathcal{{C}}(t_2,t_1) = e^{-D_\phi(t_2-t_1)} \cos \left [\frac{\Omega_0}{\gamma} (\sin(\gamma t_2) - \sin(\gamma t_1)) \right ]
\end{equation}
for $t_2>t_1$. Fig. (\ref{fig:corr_det}) shows the correlation function $\mathcal{C}(t,0)$ obtained from Langevin dynamics simulations together with the theoretical prediction Eq. (\ref{eq:corr_deterministic}). The period $t_0$ of the oscillations depends on both time scales $\gamma$ and $\Omega_0$, and can be simply extracted from Eq. (\ref{eq:corr_deterministic}), resulting in $\gamma t_0 = \sin^{-1}(2\pi \gamma/\Omega_0)$, which in the case of Fig. (\ref{fig:corr_det}) gives $t_0 \approx 1.11$. The exponential envelope is given by the rotational diffusivity as expected.

The presence of two time scales that govern the angular dynamics is also seen to have an effect on the effective diffusivity. Fig. (\ref{fig:collected} A) shows the mean-squared displacement obtained through Langevin dynamics simulations where the dynamical equations Eq. (\ref{eq:L1},\ref{eq:L2}) are integrated using a simple Euler scheme. The effective diffusivity, extracted as the linear slope at late times, displays a non-monotonic dependence on the angular speed variable $\Omega_0$, as shown in Fig. (\ref{fig:collected} B).  In Fig. (\ref{fig:collected} C) we highlight the $\gamma = 1$ case, and show typical particle trajectories corresponding to the peaks and valleys of the effective diffusivity. We see that the peaks of the plot, corresponding to enhanced transport, is due to synchronized curling events where particles in the ensemble are able to maintain their direction after tumbles, similar to what has been observed for noisy angular dynamics \cite{weber2011active}. In the valleys of the plot, the particle trajectories are more disordered. These effects are as mentioned due to temporally synchronized curling events for all trajectories in the ensemble, which may be appropriate if the torques originate from externally applied fields that affect several particles equally, or if one is interested in the typical behavior of a single long particle trajectory.  A model that includes angular motion reversal but lacks this synchronicity is discussed in the next section.

\begin{figure}[t]
    \centering
    \includegraphics[width = 8.6cm]{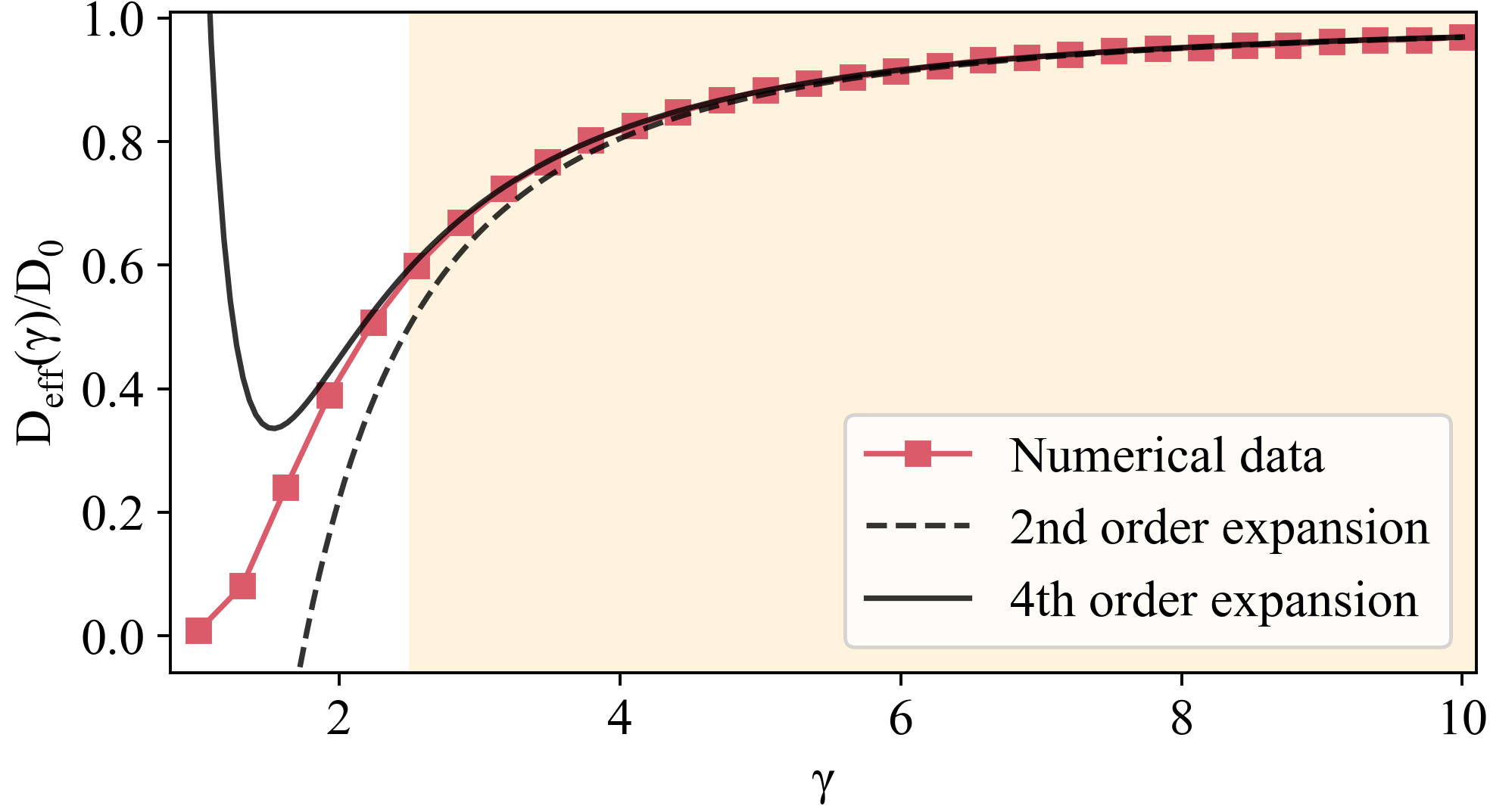}
    \caption{Effective diffusivity for deterministic reversal, showing good agreement with the prediction from the fourth order perturbative expression $D_\text{eff}^{(4)}$. Simulated with $3\cdot 10^5$ particles by numerically integrating the stochastic equations of motion. Maximal angular velocity is set to $\Omega_0 = 5/2$. Shaded region indicates the expected range of validity for the perturbative expansion. Small values of the reversal frequency demands larger numerical simulations (with simulation time being much larger than $2\pi/\gamma$), which is the reason why the behavior is not investigated in the $\gamma<1$ region. }
    \label{fig:Deff_det}
\end{figure}

Some insight regarding the dependence of the effective diffusivity on the reversal rate $\gamma$ can be obtained by analytically considering the perturbative regime where $\varepsilon \equiv \Omega_0/\gamma$ is small. This regime correspond to particles displaying meandering swimming paths without any curling. We express the effective diffusivity as a series
\begin{equation}
    D_\text{eff} = \sum_{m\geq 0} d_m(t) \varepsilon^m,
\end{equation}
where the expansion coefficients are found by Taylor expanding the cosine in the correlation function in Eq. (\ref{eq:corr_deterministic}). Since the series expansion of the cosine has only even terms, we immediately have that $d_{2m+1} = 0$ for all $m>0$. The second expansion coefficient can be calculated to be
\begin{align}
d_2  = & -\frac{3 u_0^2\gamma^3  \sin (2 t \gamma )}{2 \left(D_{\phi }^2+\gamma
   ^2\right) \left(D_{\phi }^2+4 \gamma ^2\right)}\nonumber\\\nonumber
   & -\frac{u_0^2  D_{\phi } \gamma^2
   \cos ^2(t \gamma )}{\left(D_{\phi }^2+\gamma ^2\right) \left(D_{\phi }^2+4
   \gamma ^2\right)}\\ \nonumber
   & +\frac{u_0^2\gamma ^4\cos (2 t \gamma )}{D_{\phi }
   \left(D_{\phi }^2+\gamma ^2\right) \left(D_{\phi }^2+4 \gamma
   ^2\right)}\\
  & -\frac{2u_0^2 \gamma ^4 }{D_{\phi } \left(D_{\phi }^2+\gamma
   ^2\right) \left(D_{\phi }^2+4 \gamma ^2\right)} .
\end{align}
The remaining time dependence is of an oscillatory nature, and can be dealt with by introducing the average over the time scale associated with the oscillation frequency $\overline{G} \equiv \frac{\gamma}{2 \pi} \int_0^{2 \pi /\gamma} ds G(s)$. Performing this average results in 
\begin{equation}
    \overline{d_2} =  - \frac{D_0}{2}\frac{\gamma^2}{\gamma^2 +  D_\phi^2}.
\end{equation}
This correction leads to an effective diffusivity that increases with increasing reversal rate $\gamma$. This is sensible, since we expect that in the limit of infinitely fast switching the behavior of linear swimmers should emerge. This is because at finite angular velocities, the particle will not have had time to depart from its trajectory substantially before the handedness again changes.

In a similar way one can calculate the fourth order coefficient. In this case the number of terms involved grows rather large, and one should carefully keep track of which terms will be present in the final result. Just like for the second order coefficient, one keeps only constant and oscillating terms and performs a temporal average. This results in
\begin{equation}
    \overline{d_4} = \frac{3 D_0}{8} \frac{\gamma^4}{D_\phi^4 + 5 D_\phi^2 \gamma^2 + 4 \gamma^4}.
\end{equation}
We denote the effective diffusivity containing terms up to $m$'th order $D_\text{eff}^{(m)}$. In summary, the expansion coefficients read:
\begin{align}
    &\overline{d_0} = D_0,\\
    &\overline{d_1} = 0,\\
    &\overline{d_2} =  - \frac{D_0}{2}\frac{\gamma^2}{\gamma^2 +  D_\phi^2}, \\
    &\overline{d_3} = 0,\\
    &\overline{d_4} = \frac{3 D_0}{8} \frac{\gamma^4}{D_\phi^4 + 5 D_\phi^2 \gamma^2 + 4 \gamma^4},\\
    &\overline{d_5} = 0.
\end{align}
One should note that in the above expressions, the expansion coefficients take values in $(0,1)$. This leads to a rather well-behaved perturbative series where the corrections to $D_\text{eff}^{(m)}/D_0$ in magnitude are always smaller than $\varepsilon^{m+1}$. For example, if $\varepsilon = 0.4$ the correction to the fifth order calculation provided above is already smaller than half a percent. 

Fig. (\ref{fig:Deff_det}) shows the effective diffusivity for $D_\phi = 0.1$, showing a convergence to the linear swimmer result when the reversal rate diverges. We see that transport is suppressed with  decreasing reversal rate as expected from the analytical predictions.

\section{Stochastic angular velocity reversal}
In many realistic scenarios, the switching between right- and left-handedness may not be predictable and deterministic as in the previous example. For example, the reversal rate may not be constant throughout a population of particles, or there may be random phases in the periodic behavior of the angular reversal that de-synchronize the curling events leading to the non-monotonic behavior observed in Fig. (\ref{fig:collected}B). In this section we consider a simple stochastic version of the above deterministic model.

\begin{figure}[t]
    \centering
    \includegraphics[width = 8.6cm]{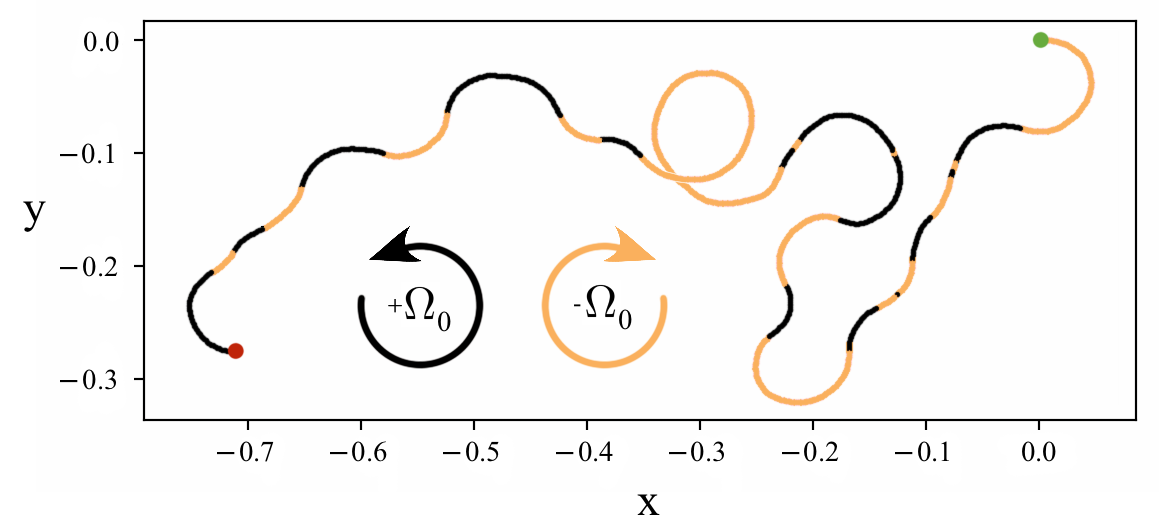}
    \caption{Sample trajectory for a particle with stochastic angular velocity reversal. Green and red points indicate initial and final position respectively.  }
    \label{fig:traj_kac}
\end{figure}

\begin{figure}[t]
    \centering
    \includegraphics[width = 8.6cm]{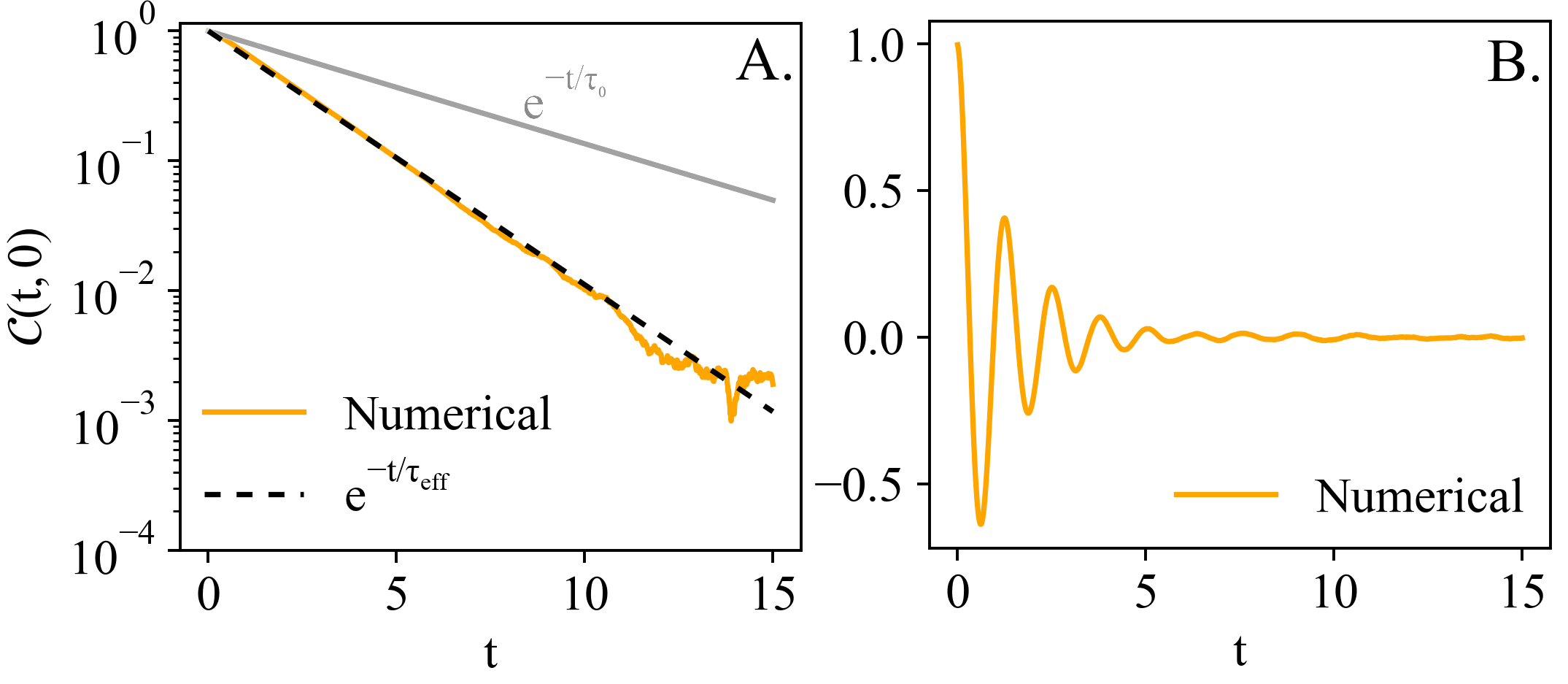}
    \caption{  Correlation function for active Brownian particles with stochastic angular velocity reversal.  A) Perturbative regime where $D_\phi = 1/5, \gamma = 2,\Omega_0 = 1, u_0 = 1$, where an exponential decay of correlations is observed. Decay with the effective persistence time scale $\tau_\text{eff}$ from Eq. (\ref{eq:timescale}) shown in dashed line. Gray line corresponds to the unperturbed persistence time.  B) Non-perturbative regime where  $D_\phi = 1/5, u_0 = 1, \gamma = 0.5,\Omega_0 = 5$, showing oscillating correlations with exponential envelope. Ensemble over $2\cdot 10^5$ particles.}
    \label{fig:Kac_corr_1}
\end{figure}

\begin{figure}[t]
    \centering
    \includegraphics[width = 8.6cm]{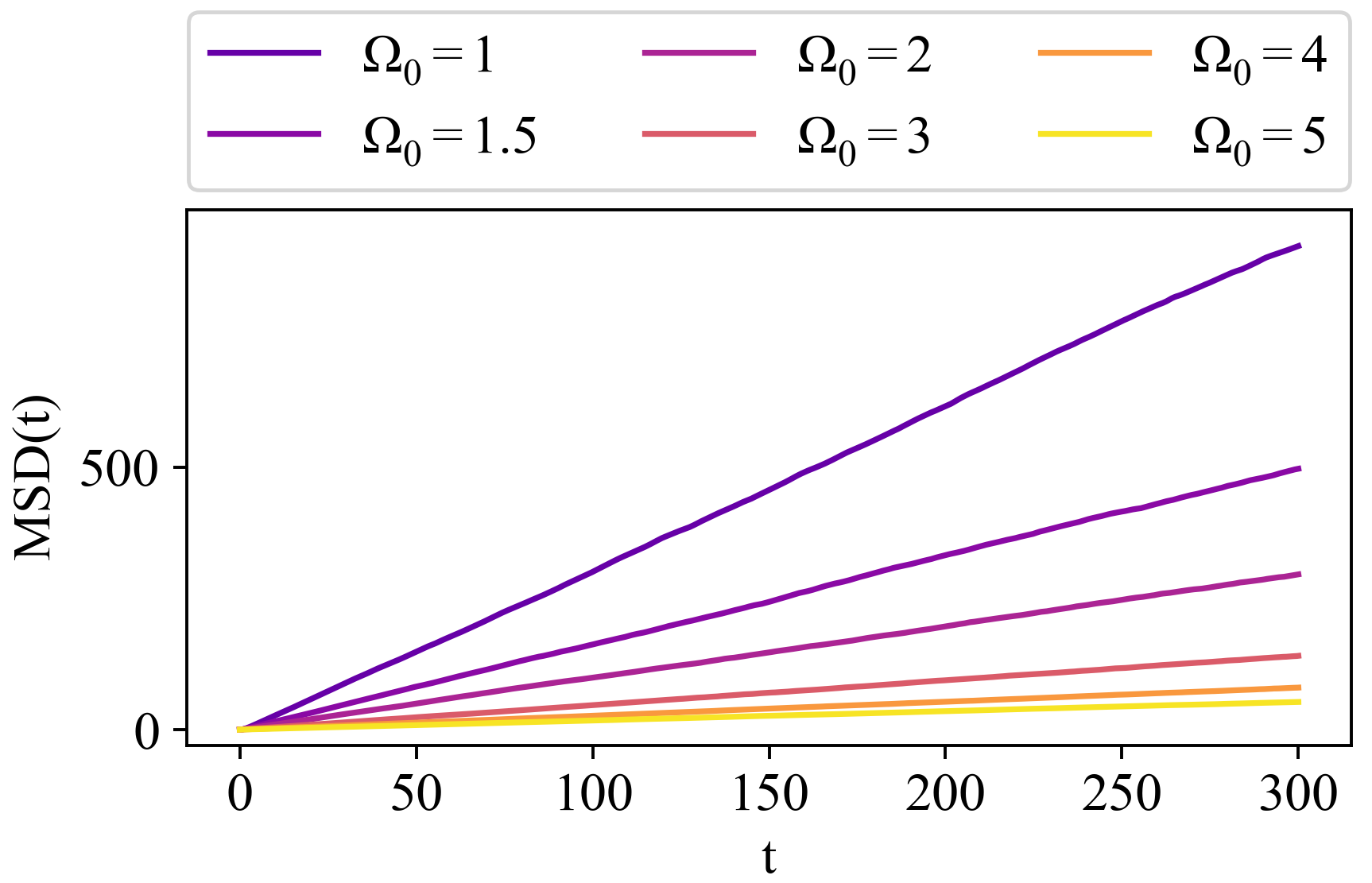}
    \caption{  Mean-squared displacements for active Brownian particles with stochastic angular velocity reversal. In contrast to the case of deterministic reversal, the diffusivity (slopes) are here decreasing monotonically with increasing maximal angular speed. Parameters used: $\Omega_0 = 5, \gamma = 1, u_0 = 1, D_\phi = 0.2$. }
    \label{fig:Kac_msds}
\end{figure}

\begin{figure}[t]
    \centering
    \includegraphics[width = 8.6cm]{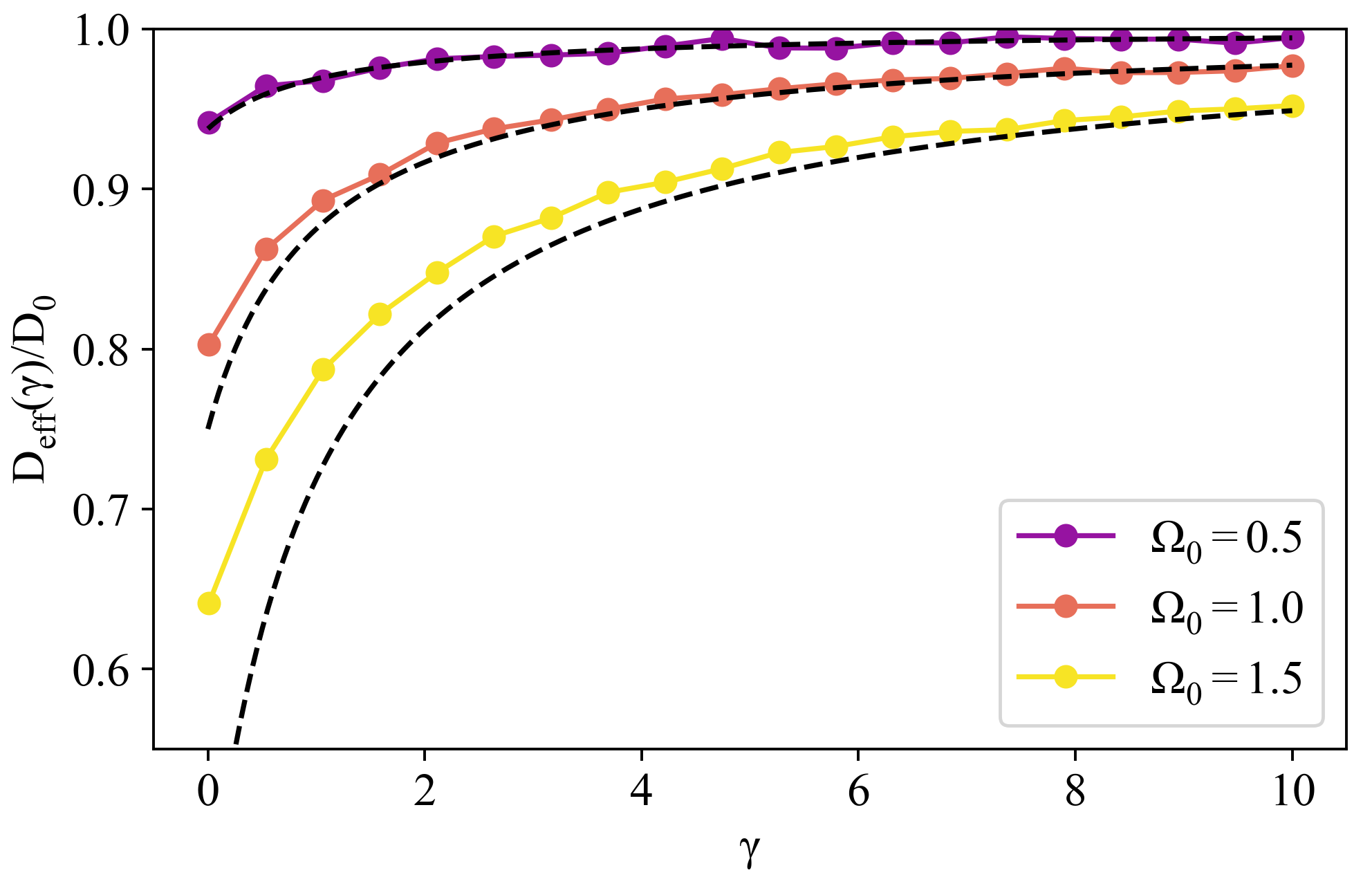}
    \caption{Effective diffusivity as a function of switching rate $\gamma$ for different values of angular speed $\Omega_0$. Colored solid lines represent numerical results obtained by performing ensemble averages over $5\cdot 10^5$ particle trajectories. Dashed lines are theoretical expression in Eq. (\ref{eq:Kac_deff_formula}).  Parameters used: $D_\phi = 2, u_0 = 1$.}
    \label{fig:Kac_deff_1}
\end{figure}

The angular velocity is now assumed to take the form $\Omega(t) = \Omega_0 \kappa(t)$, where $\kappa(t)$ is a Kac process (also known as a telegraph process). This process switches between $+1$ and $-1$ at random times $\{t_i\}$ determined by a Poisson process with rate $\gamma$. Such stochastic reversal models have been considered in the past in the context of microswimmers reversing their direction of motion \cite{santra2021active}. Fig. (\ref{fig:traj_kac}) shows a typical particle trajectory with angular velocity reversal. The mean and variance of the Kac process are known to satisfy 
\begin{align}
    &\langle\kappa(t) \rangle_\kappa = e^{-2\gamma t},\\
    &\langle\kappa(t_1)\kappa(t_2) \rangle_\kappa = e^{-2\gamma |t_2-t_1|},
\end{align}
where $\gamma$ is the reversal rate and the subscript on the angular brackets indicates that the average is taken over realizations of the $\kappa$ process. 

Proceeding similarly as in the previous section, the correlation function will for small values of the maximal angular speed compared to the time scale introduced by the switching rate take the form 
\begin{equation}\label{eq:24}
    \mathcal{C}(t_2,t_1) = e^{-D_\phi (t_2-t_1)} \left [ 1 - \frac{\Omega_0^2}{2} \mathcal{C}_\kappa (t_2,t_1)\right ],
\end{equation}
where we introduced the function
\begin{equation}
    \mathcal{C}_\kappa (t_2,t_1) = \int_{t_1}^{t_2} ds_2 \int_{t_1}^{t_2} ds_1 \langle \kappa(s_1)\kappa(s_2)\rangle_\kappa.
\end{equation}
Using the mean and variance of the reversal variable $\kappa$ one can straightforwardly show that 
\begin{equation}\label{eq:Kac_pert_corr}
    \mathcal{C}_\kappa (t_2,t_1)= \frac{t_1+t_2}{\gamma} + \frac{2}{4\gamma^2} \left (  e^{-2\gamma (t_2-t_1)} - 1 \right )
\end{equation}
for $t_2> t_1$. 

In the region of parameter space where this perturbative analysis holds, the particles are in the meandering regime where curling is unlikely to take place. In this regime one expects the correlation function to decay exponentially, and one can extract an effective persistence time scale through the formula 
\begin{equation}
    \tau^{-1}_\text{eff} = \lim_{t\to \infty} \left\{- \frac{\log \mathcal{C}(t,0)}{t} \right\}.
\end{equation}
After some algebra, the first order correction from the switching on the persistence time results in
\begin{equation}\label{eq:timescale}
\tau_\text{eff} = \frac{1}{D_\phi + \frac{\Omega_0^2}{2\gamma}}    
\end{equation}
We see that as the angular velocity vanishes one regains the linear swimmer result $\tau_0 = 1/D_\phi$. This limit is also regained as the switching frequency diverges.The switching from left- to right-handed angular motion, leading to meandering paths, will reduce the persistence time when compared to that of a linear swimmer. Fig. (\ref{fig:Kac_corr_1} A) shows the exponential decay of the correlation function, obtained numerically by calculating the ensemble average $\langle \hat P(t) \cdot \hat P(0)\rangle$ over particle trajectories. The theoretical prediction agrees well with the numerics, and deviates quite significantly from the curve corresponding to the "bare" persistence time $\tau_0$. Fig. (\ref{fig:Kac_corr_1}  B) shows the correlation function obtained numerically outside of the perturbative regime, where oscillatory behavior is observed. 

Eq. (\ref{eq:Kac_pert_corr}) together with Eq. (\ref{eq:24}) can also be used to estimate an effective diffusivity, analogous to the calculations performed in the deterministic case. To second order the effective diffusivity takes the form 
\begin{equation}\label{eq:Kac_deff_formula}
    \frac{D_\text{eff}}{D_0} = 1 - \frac{\gamma^2}{D_\phi^2 + 2 \gamma D_\phi} \left(\frac{\Omega_0}{\gamma}\right)^2
\end{equation}
Figure 7 shows the mean square displacement of active particles with stochastic reversals, obtained numerically. Fig. (\ref{fig:Kac_deff_1}) shows the effective diffusivity as a function of the switching frequency, showing good quantitative agreement in the domain of validity and good qualitative agreement also elsewhere. As in the deterministic case, the linear swimmer results are obtained as the switching frequency $\gamma$ grows very large compared to the angular speed $\Omega_0$. Lower values of switching frequency drastically reduces transport, approaching the chiral active Brownian result in the $\gamma \to 0 $ limit.

\section{Discussion}\label{sec:con}

The dynamics of self-propelled particles with meandering or curling swimming paths have been studied using numerical and analytical methods. Both deterministic and stochastic reversal of the angular velocity has been considered. In the deterministic case, a synchronization effect lead to non-monotonic behavior of the effective diffusivity as a function of maximal angular speed. The dependence of the effective diffusivity on reversal rate was studied using perturbative methods in the meandering regime, which was verified using Langevin dynamics simulations. In the case of stochastic angular velocity reversal, the effective diffusivity displays a monotonic dependence on maximal angular speed, in contrast to the deterministic case. Effective diffusivity and effective persistence times were studied analytically and verified through simulations. In particular, it was found that the effective persistence time in the regime of swimming paths with small oscillations decrease with maximal angular speed and increases with reversal rate.

\begin{acknowledgements}
%The authors thank  ...  for insightful input and discussions during this work.
The author thanks L. Angheluta, E. G. Flekkøy and I. S. Haugerud for insightful discussions. This work was supported by the Research Council of Norway through the Center of Excellence funding scheme, Project No. 262644(PoreLab). \end{acknowledgements}

\bibliographystyle{apsrev4-2}
\bibliography{OlsenChiral.bib}

\end{document}